# TEACHING THE SOCIAL MEDIA GENERATION: RETHINKING LEARNING WITHOUT SACRIFICING QUALITY

S. Azimi

*Delft University of Technology (NETHERLANDS)*

## Abstract

The rapid generational turnover and the pervasive influence of social media have reshaped how students learn, engage with teaching materials, and build trust in educational content. Today's students are accustomed to fast information processing, quick judgment, and a tendency to disengage when content does not align with their expectations. Additionally, the rise of AI tools like ChatGPT has led to growing concerns about over-reliance on technology and the lack of independent critical thinking. These shifts create significant challenges for educators, who must balance innovation with maintaining educational quality, fostering independent work, and improving course attendance. It is evident that in today's rapidly evolving digital landscape, education must adapt to meet students in their own spaces while maintaining academic integrity, quality, and core educational values.

This study explores a blended teaching approach that adapts to changing learning habits while keeping students engaged and helping them retain knowledge. It was implemented in a first-year software engineering course at a Dutch technical university. The approach includes:

(1) flipped classrooms approach, where students watch short, whiteboard animation videos before class and use class time for hands-on practice;

(2) narrative-style handouts, provided after practice to reinforce learning without replacing active participation;

(3) Compulsory first drafts of team-based formative assignments, completed in class without generative AI to ensure attendance, teamwork, and problem-solving without over-reliance on technology,

(4) weekly anonymous feedback to continuously improve content and teaching methods in real-time.

The results were highly positive: course attendance increased by 50%, with almost no dropouts. Students found the short videos helpful in reviewing material and preparing for exams. Teachers also benefited, as they could build stronger connections with students during in-class activities and better understand their challenges. The shows a clear increase in student engagement compared to previous years, leading to improved performance in both the final written test and the project.

## 1 INTRODUCTION

Today's students have grown up surrounded by smartphones, social media, and instant access to information. Platforms like TikTok, YouTube, and Instagram shape how they consume content and interact with the world [1]. This daily reality creates new challenges for university educators. Students prefer short, visual materials and quick feedback over long, traditional lectures [2,3]. They expect active involvement rather than passive listening [4-6].

However, while students are highly skilled in navigating digital spaces, they do not automatically excel at using technology for learning [7]. Studies show that students often struggle to focus when technology is constantly present. Even the silent presence of a smartphone can reduce attention and memory during learning [8, 9]. Social media use in class often leads to multitasking, which weakens understanding and lowers academic performance [10 -12].

Research suggests that effective teaching today must strike a delicate balance. It must respect how students are used to processing information but must not sacrifice depth or rigor [13]. Short videos and micro-learning activities have been shown to boost initial engagement, especially when paired with active classroom strategies [14-15]. Blended learning models, which combine online and face-to-face instruction, are up-and-coming. Well-designed blended courses often achieve better or equal outcomes than entirely traditional ones [16].

However, technology alone does not guarantee success. The purpose of any tool must be connected to learning goals. Otherwise, students risk staying busy without achieving deep understanding [17]. Guidance, structure, and clear expectations remain essential in digital and blended learning environments [18]. Instructors must design experiences that guide students beyond surface engagement toward deeper thinking and critical reflection.

These challenges have become even more urgent with the rise of generative artificial intelligence tools. While tools like AI can support learning when used thoughtfully, they also create new risks. Easy access to ready-made answers can weaken students' ability to think independently, solve problems collaboratively, and develop the deeper understanding that academic work demands. Educators must design courses that capture students' attention and reinforce habits of critical inquiry, collaboration, and thoughtful use of digital resources.

Based on this understanding, educators must rethink how they design learning experiences for the social media generation. The goal is not to make education easier or faster; it is to create learning experiences that align with how students approach information today, while still encouraging them to think critically, engage actively, and develop a real understanding. Clear expectations, strong learning structures, and careful use of technology make this possible.

This paper presents a case study where a first-year technical course at a Dutch University was redesigned to respond to these challenges. The redesigned course introduced short, animated videos to prepare students for class, focused in-class time on active teamwork and problem-solving, and carefully structured assignments to promote critical thinking without over-reliance on AI tools. Student engagement, attendance, and performance were closely monitored throughout the course.

This paper first introduces the course context, and the main challenges observed before the redesign. It then explains the blended teaching strategy that was developed. The findings show how students responded to the new course structure, both in their engagement and their performance. Building on these results, the paper reflects on what worked, what challenged students, and how future designs could support deeper learning even more effectively. The paper closes with conclusions on how higher education can better align with students' digital habits while maintaining academic rigor and fostering deeper learning.

## 2 METHODOLOGY

### 2.1. Case Context and Problem Definition

This study focuses on a first-year technical course offered at a Dutch university. The course introduces students to system development and rapid prototyping, combining theoretical foundations with hands-on practice. Most students begin with limited technical background, and adjusting to a more independent, problem-solving approach has been a consistent challenge.

In earlier editions of the course, students frequently expressed concerns about the learning approach. Many reported relying heavily on lecture slides as their primary source of information, instead of developing a deeper understanding through wider study. There was a strong tendency toward last-minute exam preparation, with students hoping that the slides and a single practice exam would be sufficient to succeed. When discrepancies arose between the practice materials and the final assessment, students felt unprepared and frustrated.

This pattern resulted in surface-level engagement and diminished learning outcomes. Students who relied on shortcuts instead of active participation found it challenging to achieve the course goals. Additionally, some began using generative AI tools for quick answers rather than solving problems independently, which further alienated them from deeper learning processes.

Recognizing these issues, the teaching team concluded that minor adjustments would not suffice. A complete redesign was necessary to foster active preparation, critical thinking, and stronger connections to the course material. The new design aimed to encourage consistent participation throughout the course, reduce reliance on last-minute study tactics, and support students in developing deeper learning habits both with and beyond digital tools.

### 2.2. Methodology: Blended Teaching Approach

To address the challenges described, the course was completely redesigned around a blended learning strategy. The most important change was the way students were introduced to the material. Instead of relying on dense slides and long lectures, the teaching team created **short, animated videos** that explained key concepts. Each video was limited to about five minutes. They were designed to match how students naturally consume content on platforms like YouTube and TikTok. Students watched these videos before coming to class. Short online quizzes followed each video to ensure that students engaged with the material and arrived prepared.

Changing the format of the educational videos alone was not enough. The real difference came during the **face-to-face classes**, where passive listening was replaced by active learning. Instructors kept in-class explanations short and immediately transitioned to small group activities. Students tackled practical problems, discussed real examples, and assessed their understanding through quick smartphone polls. They collaborated in small teams, encouraging even the quieter students to participate.

A second key innovation was introduced for the **assignments**. Students now had to complete the first draft of every major assignment during class hours. Phones, laptops, and collaboration were allowed, but no external AI tools could be used. This forced students to engage deeply with the course material instead of outsourcing their thinking. It also created a structured environment for teamwork, immediate feedback, and problem-solving. Instructors moved among groups, answering questions and offering guidance. This strategy directly responded to earlier complaints about unclear assignment expectations. By working in class under supervision, students received feedback early and avoided surprises close to deadlines.

To ensure that student voices continued to shape the course as it progressed, the teaching team introduced a **weekly anonymous feedback survey**. Each survey was short, asking students what helped their learning that week and what could be improved. Instructors reviewed the feedback carefully and made small adjustments where possible. When students struggled with a concept, additional examples or clarifications were added to the next session. When timing felt too rushed, activities were adjusted. This feedback loop built trust and helped students feel that their learning needs were taken seriously.

The blended course design intentionally mixed modern digital habits with proven educational principles. Short videos fit students' preferred media style, but the quizzes demanded real engagement. In-class activities satisfied their interaction needs but were rooted in serious learning goals. Structured teamwork tackled the problems of unclear assignments and over-reliance on AI. Continuous feedback helped prevent small frustrations from growing into larger obstacles. Every element was tied directly to the goal of improving both engagement and academic outcomes, with a strong commitment to maintaining the course's rigor.

## 3  RESULTS

The redesigned course resulted in a clear and measurable improvement. Attendance rose by nearly fifty percent compared to the previous year. Classrooms, once half empty, became filled with students actively engaging with the material. This change in attendance patterns significantly influenced the learning outcomes.

Students responded positively to the new blended structure. The short, animated videos were valued as a helpful tool for reviewing and preparing before class. However, students quickly realized that the videos alone were not enough to fully understand the course material. They reported that in-class activities, group discussions, and problem-solving sessions were essential for mastering the content.

This shift in engagement was reflected in the final results. None of the students who attended class regularly failed the final exam. Students who consistently participated in lectures and working sessions achieved higher overall grades than those who attended less often. The connection between active participation and academic success became clear by the end of the course.

While the majority of students appreciated the new format, not all reactions were purely positive. Some students expressed frustration that attending lectures and staying active in class was necessary to succeed. They had hoped that watching the videos might be enough. This expectation clashed with the course design, which demanded active involvement to reach the deeper levels of understanding required for the final assessment.

Feedback on the lectures themselves was encouraging. Students described the sessions as clear, focused, and supportive of their learning. They valued the structured working lectures where they could apply theories through hands-on tasks. Students especially highlighted the role of teaching assistants during these sessions, noting that the opportunity for direct help and discussion made a real difference in their ability to grasp difficult topics.

Many students adapted quickly to the new approach, recognizing that steady participation was key to their success. The pattern was clear by the end of the course: students who made use of both the videos and the active sessions performed better and felt more confident in their understanding. Although a few students wished that the videos alone would be enough, most came to see that real learning happened through discussion, practice, and feedback in class. The experience confirmed that matching the format of a course to students' digital habits can encourage deeper learning, as long as it is paired with clear expectations and serious academic challenges.

## 4 CONCLUSIONS

The results of the redesigned course show that small, thoughtful changes in structure can have a large impact. Students engaged more when given material in a format they recognized, yet real learning still happened during active in-class sessions. Short videos helped students prepare, but it was through discussion, practice, and application that they built a deeper understanding.

The link between class attendance and performance was especially clear. Students who attended regularly did not fail the final exam. Those who participated steadily achieved higher grades overall. These outcomes underline that while digital tools offer valuable flexibility, meaningful learning still depends on human interaction, guided practice, and structured support.

Managing the role of AI tools in the course also brought important lessons. Restricting AI use during early assignment work encouraged students to develop their own ideas first, while allowing later refinement recognized that technology would be part of their future practice. Some students, however, expressed a wish for more structured guidance on how to use AI responsibly. Future versions of the course could integrate this dimension more openly, helping students learn not just when to use technology, but how to do so thoughtfully.

The shift in expectations also challenged some students. A few expressed frustrations that success depended on active attendance and participation. This reflects a broader tension in modern education, where students sometimes expect learning to fit fully into their digital habits. The experience of this course suggests that blending digital preparation with strong in-person work can meet students halfway, but not fully replace the need for time, effort, and presence.

While the redesigned course clearly succeeded in improving engagement and learning outcomes, there are limits to what this study can claim. It focused on a single course in a single year, and results might differ in other subjects, institutions, or student groups. Even when digital methods fit students' habits, clear structure, consistent communication, and active challenges remain essential.

Finally, the experience highlights a hopeful path forward. Technology can be an ally, but it cannot replace the hard work of real learning. When educators combine digital familiarity with high expectations, clear guidance, and active learning, students are willing to rise to the challenge — and often surpass it.